\documentclass{article}
\usepackage{graphicx}
\usepackage{amsmath}
\usepackage{amsfonts}
\usepackage{amssymb}

\input epsf

\begin{document}

\title{A non-perturbative analysis of symmetry breaking in two-dimensional $\phi^{4}$
theory using periodic field methods}
\author{Pablo J. Marrero\thanks{email: marrero@het.phast.umass.edu}, Erick A.
Roura\thanks{email: roura@het.phast.umass.edu}, and Dean Lee\thanks{email: dlee@het.phast.umass.edu}\\Department of Physics and Astronomy\\University of Massachusetts, Amherst, MA \ 01003}
\maketitle
\begin{abstract}
We describe the generalization of spherical field theory to other modal
expansion methods. The main approach remains the same, to reduce a
$d$-dimensional field theory into a set of coupled one-dimensional systems.
\ The method we discuss here uses an expansion with respect to periodic-box
modes. \ We apply the method to $\phi^{4}$ theory in two dimensions and
compute the critical coupling and critical exponents. \ We compare with
lattice results and predictions via universality and the two-dimensional Ising
model. \ [PACS numbers: 05.10-a, 05.50+q, 11.10.-z, 11.30.Qc, 12.38Lg]
\end{abstract}

\section{Introduction}

Recently spherical field theory has been introduced as a non-perturbative
method for studying quantum field theory \cite{spher}. \ The starting point of
this approach is to decompose field configurations in a $d$-dimensional
Euclidean functional integral as linear combinations of spherical partial
waves. Regarding each partial wave as a distinct field in a new
one-dimensional system, the functional integral is rewritten as a
time-evolution equation, with radial distance serving as the parameter of
time. \ The core idea of spherical field theory is to reduce quantum field
theory to a set of coupled quantum-mechanical systems. \ The technique used is
partial wave decomposition, but this can easily be generalized to other modal
expansions. \ Instead of concentric spheres and partial waves, we might
instead consider a generic smooth one-parameter family of ($d-1$)-dimensional
manifolds (disjoint and compact) and basis functions defined over each
manifold. \ There are clearly many possibilities and we can optimize our
expansion scheme to suit the specific problem at hand.

One important and convenient feature of spherical field theory is that it
eliminates the need to compactify space --- the spherical quantization surface
is already compact. \ This is useful for studying phenomena which might be
influenced by our choice of boundary conditions, such as topological
excitations. \ Another interesting example arises in the process of
quantization on noncommutative geometries. \ Spherical field methods have
recently been adapted to study quantum field theory on the noncommutative
plane \cite{chaichian}. \ In many instances, however, maintaining exact
translational invariance is of greater value than non-compactness or exact
rotational invariance. \ In that case one could consider a system whose
spatial dimensions have been compactified to form a periodic box. \ The next
step would be to expand field configurations using free modes of the box and
evolve in time. We will refer to this arrangement as periodic-mode field
theory or, more simply, periodic field theory.\footnote{Periodic field theory
could be viewed as a hybrid of the Hamiltonian and momentum-space lattice
formalisms. \ This combination, however, has not been discussed in the literature.}

In periodic field theory the Hamiltonian is time independent and linear
momentum is exactly conserved. \ In this work we describe the basic features
of periodic field theory and use it to analyze spontaneous symmetry breaking
in Euclidean two-dimensional $\phi^{4}$ theory. \ We use the method of
diffusion Monte Carlo to simulate the dynamics of the theory. \ The techniques
discussed here have several advantages over conventional Euclidean lattice
Monte Carlo methods. \ One is that periodic boundary conditions are not
imposed on the time variable, making it easier to determine the particle mass
from an exponential decay fit. \ Another is that the zero-momentum mode is
regarded as a single degree of freedom (rather than a collective mode on the
lattice), which provides a simpler description of vacuum expectation values
and symmetry breaking. \ Other advantages arise in systems not considered
here, such as the absence of fermion doubling and extensions to Minkowski
space using non-stochastic computational methods.

The organization of this paper is as follows. We begin with a derivation of
periodic field theory. We then analyze two-dimensional $\phi^{4}$ theory and
its corresponding periodic-field Hamiltonian using diffusion Monte Carlo
methods. \ We compute the critical coupling at which $\phi\rightarrow-\phi$
reflection symmetry is broken, and determine the critical exponents $\nu$ and
$\beta$.\footnote{We are using standard notation. \ $\nu$ is associated with
the inverse correlation length, and $\beta$ corresponds with the behavior of
the vacuum expectation value of $\phi$.} \ We find that our value of the
critical coupling is in agreement with recent lattice results \cite{loinaz},
and our values for the critical exponents are consistent with\ predictions via
universality and the two-dimensional Ising model.

\section{Periodic fields}

We start with free scalar field theory in two dimensions subject to periodic
boundary conditions
\begin{equation}
\phi(t,x-L)=\phi(t,x+L).
\end{equation}
In our discussion $t$ is Euclidean time, analytically continued from imaginary
Minkowskian time. Let $\mathcal{J}$ be an external source satisfying the same
boundary conditions and which vanishes as $\left|  t\right|  \rightarrow
\infty$. \ The Euclidean generating functional in the presence of
$\mathcal{J}$ is given by
\begin{equation}
Z[\mathcal{J}]\propto\int\mathcal{D}\phi\exp\left\{  -%
{\textstyle\int_{-\infty}^{\infty}}
dt%
{\textstyle\int_{-L}^{L}}
dx\left[  \tfrac{1}{2}\left(  (\tfrac{\partial\phi}{\partial t})^{2}%
+(\tfrac{\partial\phi}{\partial x})^{2}\right)  +\tfrac{\mu^{2}}{2}\phi
^{2}-\mathcal{J}\phi\right]  \right\}  .
\end{equation}
We now expand in terms of periodic-box modes,
\begin{align}
\phi(t,x) &  =\sqrt{\tfrac{1}{2L}}%
{\textstyle\sum_{n=0,\pm1,...}}
\phi_{n}(t)e^{in\pi x/L},\\
\mathcal{J}(t,x) &  =\sqrt{\tfrac{1}{2L}}%
{\textstyle\sum_{n=0,\pm1,...}}
\mathcal{J}_{n}(t)e^{in\pi x/L}.\nonumber
\end{align}
These are also eigenmodes of momentum and each $\phi_{n}$ or $\mathcal{J}_{n}$
carries momentum $\frac{n\pi}{L}$. \ In terms of these modes, we have
\begin{equation}
Z[\mathcal{J}]\propto\int%
{\textstyle\prod_{n}}
\mathcal{D}\phi_{n}\exp\left\{  -%
{\textstyle\int_{-\infty}^{\infty}}
dt%
{\textstyle\sum_{n}}
\left[  \tfrac{1}{2}\tfrac{d\phi_{-n}}{dt}\tfrac{d\phi_{n}}{dt}+\tfrac{1}%
{2}\left(  \tfrac{n^{2}\pi^{2}}{L^{2}}+\mu^{2}\right)  \phi_{-n}\phi
_{n}-\mathcal{J}_{-n}\phi_{n}\right]  \right\}  .
\end{equation}
For notational purposes we will define
\begin{equation}
\omega_{n}=\sqrt{\tfrac{n^{2}\pi^{2}}{L^{2}}+\mu^{2}}.
\end{equation}
Using the Feynman-Kac formula, we find
\begin{equation}
Z[\mathcal{J}]\propto\left\langle 0\right|  T\exp\left\{  -%
{\textstyle\int_{-\infty}^{\infty}}
dt\,H_{\mathcal{J}}\right\}  \left|  0\right\rangle ,
\end{equation}
where
\begin{equation}
H_{\mathcal{J}}=%
{\textstyle\sum_{n}}
\left[  -\tfrac{1}{2}\tfrac{\partial}{\partial q_{-n}}\tfrac{\partial
}{\partial q_{n}}+\tfrac{1}{2}\omega_{n}^{2}q_{-n}q_{n}-\mathcal{J}_{-n}%
q_{n}\right]  ,
\end{equation}
and $\left|  0\right\rangle $ is the ground state of $H_{0}$. \ Since $H_{0}$
is the usual equal time Hamiltonian, $\left|  0\right\rangle $ is the vacuum.
\ $H_{0}$ consists of a set of decoupled harmonic oscillators, and it is
straightforward to calculate the two-point correlation functions, \
\begin{equation}
\left\langle 0|\phi_{-n}(t_{2})\phi_{n}(t_{1})|0\right\rangle =\left.
\tfrac{\delta}{\delta\mathcal{J}_{n}(t_{2})}\tfrac{\delta}{\delta
\mathcal{J}_{-n}(t_{1})}Z[\mathcal{J}]\right|  _{\mathcal{J}=0}=\tfrac
{1}{2\omega_{n}}\exp\left[  -\omega_{n}\left|  t_{2}-t_{1}\right|  \right]
.\label{corr}%
\end{equation}

We now include a $\phi^{4}$ interaction term as well as a counterterm
Hamiltonian, which we denote as $H_{c.t.}$. \ The new Hamiltonian is
\begin{align}
H_{\mathcal{J}}  &  =%
{\textstyle\sum_{n}}
\left[  -\tfrac{1}{2}\tfrac{\partial}{\partial q_{-n}}\tfrac{\partial
}{\partial q_{n}}+\tfrac{1}{2}\omega_{n}^{2}q_{-n}q_{n}-\mathcal{J}_{-n}%
q_{n}\right] \label{hamilton}\\
&  +\tfrac{\lambda}{4!2L}%
{\textstyle\sum_{n_{1}+n_{2}+n_{3}+n_{4}=0}}
q_{n_{1}}q_{n_{2}}q_{n_{3}}q_{n_{4}}+H_{c.t.}.\nonumber
\end{align}
We will regulate the sums over momentum modes by choosing some large positive
number $N_{\max}$ and throwing out all high-momentum modes $q_{n}$ such that
$\left|  n\right|  >N_{\max}$. This corresponds to a momentum cutoff
\begin{equation}
\Lambda^{2}=\left(  \tfrac{N_{\max}\pi}{L}\right)  ^{2}.
\end{equation}
In two-dimensional $\phi^{4}$ theory, renormalization can be implemented by
normal ordering the $\phi^{4}$ interaction term. This corresponds to
cancelling diagrams of the type shown in Figure 1. Using (\ref{corr}), we
find
\begin{equation}
H_{c.t.}=\tfrac{6\lambda b}{4!2L}%
{\textstyle\sum_{n=-N_{\max},N_{\max}}}
q_{-n}q_{n},
\end{equation}
where
\begin{equation}
b=%
{\textstyle\sum_{n=-N_{\max},N_{\max}}}
\tfrac{1}{2\omega_{n}}.
\end{equation}
For the remainder of our discussion we will use the Hamiltonian%
\begin{align}
H  &  =H_{0}=%
{\textstyle\sum_{n}}
\left[  -\tfrac{1}{2}\tfrac{\partial}{\partial q_{-n}}\tfrac{\partial
}{\partial q_{n}}+\tfrac{1}{2}(\omega_{n}^{2}-\tfrac{\lambda b}{4L}%
)q_{-n}q_{n}\right] \\
&  +\tfrac{\lambda}{4!2L}%
{\textstyle\sum_{n_{1}+n_{2}+n_{3}+n_{4}=0}}
q_{n_{1}}q_{n_{2}}q_{n_{3}}q_{n_{4}}.\nonumber
\end{align}

\section{The $\phi_{2}^{4}$ phase transition}

The existence of a second-order phase transition in two-dimensional $\phi^{4}$
theory has been derived in the literature \cite{glimm}\cite{chang}. The phase
transition is due to $\phi$ developing a non-zero expectation value and the
resultant spontaneous breaking of $\phi\rightarrow-\phi$ reflection symmetry.
It is generally believed that this theory belongs to the same universality
class as the two-dimensional Ising model and therefore shares the same
critical exponents. \ In this section we apply periodic field methods to
$\phi^{4}$ theory in order to determine the critical coupling and critical
exponents $\nu$ and $\beta$. \ We have chosen $\nu$ and $\beta$ since these
are, in our opinion, the easiest to determine from direct computations. \ All
other exponents can be derived from these using well-known scaling laws.
\ From the Ising model predictions we expect
\begin{equation}
\nu=1,\quad\beta=\tfrac{1}{8}.
\end{equation}

$\nu$ is the exponent associated with the inverse correlation length or,
equivalently, the mass of the one-particle state. \ We will determine the
behavior of the mass as we approach the critical point from the symmetric
phase of the theory. \ Let $\left|  a\right\rangle $ be any state even under
reflection symmetry. \ We consider the matrix element%
\begin{equation}
f(t)=\left\langle a\right|  q_{0}\exp\left(  -tH\right)  q_{0}\left|
a\right\rangle \text{.}%
\end{equation}
Inserting energy eigenstates $\left|  i\right\rangle $ satisfying $H\left|
i\right\rangle =E_{i}\left|  i\right\rangle $, we have\
\begin{equation}
f(t)=%
{\textstyle\sum_{i}}
\exp\left(  -tE_{i}\right)  \left|  \left\langle i\right|  q_{0}\left|
a\right\rangle \right|  ^{2}. \label{sum}%
\end{equation}
Since $\left|  a\right\rangle $ and $\left|  0\right\rangle $ are even under
reflection symmetry and $q_{0}$ is odd, the vacuum contribution to the sum in
(\ref{sum}) vanishes. \ In the limit $t\rightarrow\infty,$ (\ref{sum}) is
dominated by the contribution of the next lowest energy state, the
one-particle state at rest.\footnote{We are assuming that this contribution
does not also vanish. This is generally true, and we can always vary $\left|
a\right\rangle $ to make it so.} In this limit we have%
\begin{equation}
f(t)\sim e^{-mt},
\end{equation}
where $m$ is the mass. \ We can compute $f(t)$ numerically using the method of
diffusion Monte Carlo (DMC). \ The idea of DMC is to model the dynamics of the
imaginary-time Schr\"{o}dinger equation using the diffusion and
decay/production of simulated particles. The kinetic energy term in the
Hamiltonian determines the diffusion rate of the particles (usually called
replicas) and the potential energy term determines the local decay/production
rate. A self-contained introduction to DMC can be found in \cite{kosztin}.

$\nu$ is defined by the behavior of $m$ near the critical coupling
$\lambda_{c},$%
\begin{equation}
m\sim(\lambda_{c}-\lambda)^{\nu}.
\end{equation}
\ Once we determine $f(t)$ using DMC simulations we can extract $m$ and $\nu$
using curve-fitting techniques. \ We have calculated $m$ as a function of
$\lambda$ for several different values of $L$ and $N_{\max}$. \ Results from
these calculations are presented in the next section.

$\beta$ is the critical exponent describing the behavior of the vacuum
expectation value. \ In the symmetric phase the vacuum state is unique and
invariant under the reflection transformation $\phi\rightarrow-\phi$ (or
equivalently $q_{n}\rightarrow-q_{n},$ for each $n$)$.$ \ In the
broken-symmetry phase the vacuum is degenerate as $L\rightarrow\infty$,
and\ $q_{0}$, the zero-momentum mode, develops a vacuum expectation value.
\ In the $L\rightarrow\infty$ limit tunnelling between vacuum states is
forbidden. One ground state, $\left|  0^{+}\right\rangle ,$ is non-zero only
for values $q_{0}>0$ and the other, $\left|  0^{-}\right\rangle $, is non-zero
only for $q_{0}<0$. \ We will choose $\left|  0^{+}\right\rangle $ and
$\left|  0^{-}\right\rangle $ to be unit normalized.

Let us now define the symmetric and antisymmetric combinations,%
\begin{align}
\left|  0^{s}\right\rangle  &  =\tfrac{1}{\sqrt{2}}\left(  \left|
0^{+}\right\rangle +\left|  0^{-}\right\rangle \right) \\
\left|  0^{a}\right\rangle  &  =\tfrac{1}{\sqrt{2}}\left(  \left|
0^{+}\right\rangle -\left|  0^{-}\right\rangle \right)  .\nonumber
\end{align}
We will select the relative phases of $\left|  0^{-}\right\rangle $ and
$\left|  0^{+}\right\rangle $ so that they transform from one to the other
under reflection symmetry. $\ \left|  0^{s}\right\rangle $ and $\left|
0^{a}\right\rangle $ are then symmetric and antisymmetric (respectively) under
reflection symmetry.

To avoid notational confusion in the following, we will write $\hat{q}_{n}$ to
denote the quantum-mechanical position operator and $q_{n}$ for the
corresponding ordinary variable. \ Let $\left\{  \mathcal{\hat{P}}%
_{z}\right\}  _{z\in(-\infty,\infty)}\ $be the spectral family associated with
the operator $\frac{\hat{q}_{0}}{\sqrt{2L}}$.\footnote{The extra factor
$\frac{1}{\sqrt{2L}}$ has been included for later convenience.} This implies
that $%
{\textstyle\int_{a}^{b}}
d\mathcal{\hat{P}}_{z}$ is a projection operator whose action on a general
wavefunction $\Psi$ is
\begin{equation}
\left(
{\textstyle\int_{a}^{b}}
d\mathcal{\hat{P}}_{z}\right)  \Psi(q_{0},q_{1},\cdots)=\theta(\tfrac{q_{0}%
}{\sqrt{2L}}-a)\theta(b-\tfrac{q_{0}}{\sqrt{2L}})\Psi(q_{0},q_{1},\cdots).
\end{equation}
From the support properties of $\left|  0^{+}\right\rangle $ and $\left|
0^{-}\right\rangle $, we deduce%
\begin{align}
\left|  0^{+}\right\rangle  &  =\sqrt{2}%
{\textstyle\int_{0}^{\infty}}
d\mathcal{\hat{P}}_{z}\left|  0^{s}\right\rangle =\sqrt{2}%
{\textstyle\int_{0}^{\infty}}
d\mathcal{\hat{P}}_{z}\left|  0^{a}\right\rangle \label{convert}\\
\left|  0^{-}\right\rangle  &  =\sqrt{2}%
{\textstyle\int_{-\infty}^{0}}
d\mathcal{\hat{P}}_{z}\left|  0^{s}\right\rangle =-\sqrt{2}%
{\textstyle\int_{-\infty}^{0}}
d\mathcal{\hat{P}}_{z}\left|  0^{a}\right\rangle .\nonumber
\end{align}
Using our new spectral language, we can write
\begin{equation}
\tfrac{\hat{q}_{0}}{\sqrt{2L}}=%
{\textstyle\int_{-\infty}^{\infty}}
\,z\,d\mathcal{\hat{P}}_{z}. \label{spec}%
\end{equation}

We now consider the vacuum expectation value $\left\langle 0^{+}\right|
\phi\left|  0^{+}\right\rangle $.\footnote{We could also consider
$\left\langle 0^{-}\right|  \phi\left|  0^{-}\right\rangle $. \ By reflection
symmetry $\left\langle 0^{-}\right|  \phi\left|  0^{-}\right\rangle
=-\left\langle 0^{+}\right|  \phi\left|  0^{+}\right\rangle .$} \ Making use
of translational invariance, we have
\begin{equation}
\left\langle 0^{+}\right|  \phi\left|  0^{+}\right\rangle =\tfrac{1}{2L}%
{\textstyle\int_{-L}^{L}}
dx\,\left\langle 0^{+}\right|  \phi(t,x)\left|  0^{+}\right\rangle
=\left\langle 0^{+}\right|  \tfrac{\hat{q}_{0}}{\sqrt{2L}}\left|
0^{+}\right\rangle .
\end{equation}
From (\ref{convert}) and (\ref{spec}), we conclude that
\begin{equation}
\left\langle 0^{+}\right|  \phi\left|  0^{+}\right\rangle =2\left\langle
0^{s}\right|
{\textstyle\int_{0}^{\infty}}
zd\mathcal{\hat{P}}_{z}\left|  0^{s}\right\rangle =2%
{\textstyle\int_{0}^{\infty}}
dz\,zg(z),
\end{equation}
where%
\begin{equation}
g(z)=\left\langle 0^{s}\right|  \tfrac{d\mathcal{\hat{P}}_{z}}{dz}\left|
0^{s}\right\rangle .
\end{equation}
$g(z)$ satisfies the normalization condition
\begin{equation}%
{\textstyle\int_{-\infty}^{\infty}}
dz\,g(z)=2%
{\textstyle\int_{0}^{\infty}}
dz\,g(z)=1.
\end{equation}

In our calculations we will be working with large but finite $L$. \ In this
case the ground state degeneracy is not exact, and the symmetric state
$\left|  0^{s}\right\rangle $ is slightly lower in energy than the
antisymmetric state $\left|  0^{a}\right\rangle $. \ We can now use this
observation (that $\left|  0^{s}\right\rangle $ is the lowest energy state) to
rewrite $g(z)$ as \
\begin{equation}
g(z)=\left\langle 0^{s}\right|  \tfrac{d\mathcal{\hat{P}}_{z}}{dz}\left|
0^{s}\right\rangle =\lim_{t\rightarrow\infty}\tfrac{\left\langle b\right|
\exp\left\{  -t\,H\right\}  \tfrac{d\mathcal{\hat{P}}_{z}}{dz}\exp\left\{
-tH\right\}  \left|  b\right\rangle }{\left\langle b\right|  T\exp\left\{
-2t\,H\right\}  \left|  b\right\rangle }, \label{f}%
\end{equation}
where $\left|  b\right\rangle $ is any state such that $\left\langle
0^{s}|b\right\rangle $ is non-zero.

For free field theory $g(z)$ can be exactly calculated,%
\begin{equation}
g(z)=\sqrt{\tfrac{2\mu L}{\pi}}e^{-2\mu Lz^{2}}. \label{free}%
\end{equation}
For non-trivial coupling we can calculate the right-hand side of (\ref{f})
using DMC methods. \ In Figure 2 we have plotted $g(z)$ for $L=2.5\pi$ and
$L=5\pi$. \ In each case $\tfrac{\lambda}{4!}=2.76$ and $\Lambda=4.$ \ All
quantities are measured in units where $\mu=1$. \ As can be seen, the
distributions are bimodal and the maxima for both curves occur near $\pm0.55$.
We observe that the peaks are taller and narrower for larger $L.$ This is
consistent with our intuitive picture of fluctuations in the functional
integral. \ For a small but fixed deviation in the average value of $\phi$,
the net change in an extensive quantity such as the action or total energy
scales proportionally with the size of system. \ Consequently the average size
of the fluctuations must decrease with $L$. \ We can estimate the amplitude of
the fluctuations, $\Delta\phi$, by assuming a quadratic dependence in
$\Delta\phi$ about the local minimum. \ The net effect of the fluctuation
should not scale with $L$, and we conclude that\footnote{The dimension of time
does not enter here since we are considering properties of the vacuum, the
ground state of the Hamiltonian defined at a given time. \ This is in contrast
with lattice calculations which usually consider the quantity $\left\langle
\phi\right\rangle $, the average of $\phi$ over all space and time.}%
\begin{equation}
\Delta\phi\thicksim\left(  \tfrac{1}{\sqrt{L}}\right)  ^{\#\text{spatial
dim.}}=\tfrac{1}{\sqrt{L}}. \label{fluct}%
\end{equation}
This agrees with the free field result in (\ref{free}) and also appears to be
consistent with the peak widths plotted in Figure 2.

Let $z_{\max}$ be the location of the non-negative maximum of $g(z)$. Since
$g(z)$ becomes sharply peaked as $L\rightarrow\infty$,
\begin{equation}
\left\langle 0^{+}\right|  \phi\left|  0^{+}\right\rangle =2%
{\textstyle\int_{0}^{\infty}}
dz\,zg(z)\underset{L\rightarrow\infty}{\approx}2z_{\max}%
{\textstyle\int_{0}^{\infty}}
dz\,g(z)=z_{\max}\text{.}%
\end{equation}
This gives us another option for calculating the vacuum expectation value.
\ We can either integrate 2z$\,g(z)$ or read the location of the maximal point
$z_{\max}$. \ Both will converge to the same value as $L\rightarrow\infty$.
\ However, the $z_{\max}$ result is less prone to systematic error generated
by the $O(\tfrac{1}{\sqrt{L}})$ fluctuations described above.\footnote{We can
see this explicitly in free field theory, where the vacuum expectation value
should vanish. \ $z_{\max}=0$ as desired, but $2%
{\textstyle\int_{0}^{\infty}}
dz\,zg(z)=\tfrac{1}{\sqrt{2\pi\mu L}}.$} We will therefore use%
\begin{equation}
\left\langle 0^{+}\right|  \phi\left|  0^{+}\right\rangle =z_{\max}.
\end{equation}

$\beta$ is defined by the behavior of the vacuum expectation value as we
approach the critical coupling,%
\begin{equation}
\left\langle 0^{+}\right|  \phi\left|  0^{+}\right\rangle \sim(\lambda
-\lambda_{c})^{\beta}.
\end{equation}
Using DMC methods$,$ we have computed the $\lambda$ dependence of the vacuum
expectation value for several values of $L$ and $N_{\max}.$ \ The results are
shown in the next section.

\section{Results}

The results of our diffusion Monte Carlo simulations are presented here. \ For
each set of parameters $L$ and $N_{\max},$ the curves for $m$ and
$\left\langle 0^{+}\right|  \phi\left|  0^{+}\right\rangle $ near the critical
coupling have been fitted using the parameterized forms%

\begin{equation}
m=a\left(  \tfrac{\lambda_{c}^{m}}{4!}-\tfrac{\lambda}{4!}\right)  ^{\nu}%
\end{equation}
and%
\begin{equation}
\left\langle 0^{+}\right|  \phi\left|  0^{+}\right\rangle =b\left(
\tfrac{\lambda}{4!}-\tfrac{\lambda_{c}^{\left\langle \phi\right\rangle }}%
{4!}\right)  ^{\beta}.
\end{equation}
For the mass curves, data points with $m\lesssim L^{-1}$ have correlation
lengths exceeding the size of our system and are of questionable significance.
\ We have therefore fit these curves two different ways, once using all data
points and a second time excluding small $m$ values. \ The curves for the data
set $L=2.5\pi$ and $N_{\max}=10$ are shown in Figures 3 and 4.\footnote{As
mentioned before, we are using units where $\mu=1$.} \ The error bars
represent an estimate of the error due to Monte Carlo statistical fluctuations
and, in the case of the mass data, pollution due to higher energy states.
\ Let $\tilde{\chi}_{d}^{2}$ denote the reduced chi-squared value for $d$
degrees of freedom. \ The results of the fits are as follows:

\noindent$L=2.5\pi,$ $N_{\max}=8$ ($\Lambda=3.2$):%
\begin{align}
\tfrac{\lambda_{c}^{m}}{4!}  &  =3.1\pm0.2,\ \nu=1.1\pm0.1,\ a=0.31\pm
0.02,\ \tilde{\chi}_{15}^{2}=0.81\text{ (all),}\\
\tfrac{\lambda_{c}^{m}}{4!}  &  =3.1\pm0.2,\ \nu=1.0\pm0.1,\ a=0.30\pm
0.03,\ \tilde{\chi}_{8}^{2}=0.96\ \text{(partial),}\nonumber\\
\ \tfrac{\lambda_{c}^{\left\langle \phi\right\rangle }}{4!}  &  =2.5\pm
0.1,\ \beta=0.15\pm0.02,\ b=0.65\pm0.05,\ \tilde{\chi}_{8}^{2}=0.45\nonumber
\end{align}
$L=2.5\pi,$ $N_{\max}=10,$ ($\Lambda=4$):%
\begin{align}
\tfrac{\lambda_{c}^{m}}{4!}  &  =2.9\pm0.2,\ \nu=1.1\pm0.1,\ a=0.35\pm
0.02,\ \tilde{\chi}_{12}^{2}=0.46\text{ (all),}\\
\tfrac{\lambda_{c}^{m}}{4!}  &  =2.9\pm0.2,\ \nu=1.0\pm0.1,\ a=0.37\pm
0.03,\ \tilde{\chi}_{6}^{2}=0.37\ \text{(partial),}\nonumber\\
\tfrac{\lambda_{c}^{\left\langle \phi\right\rangle }}{4!}  &  =2.5\pm
0.1,\ \beta=0.18\pm0.01,\ b=0.70\pm0.03,\ \tilde{\chi}_{9}^{2}=0.29\nonumber
\end{align}
$L=5\pi,$ $N_{\max}=16,$ ($\Lambda=3.2$):%
\begin{align}
\tfrac{\lambda_{c}^{m}}{4!}  &  =3.0\pm0.2,\ \nu=1.2\pm0.1,\ a=0.32\pm
0.03,\ \tilde{\chi}_{13}^{2}=0.87\text{ (all),}\\
\tfrac{\lambda_{c}^{m}}{4!}  &  =3.1\pm0.2,\ \nu=1.2\pm0.1,\ a=0.30\pm
0.03,\ \tilde{\chi}_{10}^{2}=0.88\ \text{(partial),}\nonumber\\
\tfrac{\lambda_{c}^{\left\langle \phi\right\rangle }}{4!}  &  =2.7\pm
0.1,\ \beta=0.12\pm0.02,\ b=0.62\pm0.04,\ \tilde{\chi}_{6}^{2}=0.50\nonumber
\end{align}
$L=5\pi,$ $N_{\max}=20,$ ($\Lambda=4$):%
\begin{align}
\tfrac{\lambda_{c}^{m}}{4!}  &  =2.8\pm0.2,\ \nu=1.2\pm0.1,\ a=0.35\pm
0.06,\text{ }\tilde{\chi}_{11}^{2}=1.2\ \text{(all),}\\
\tfrac{\lambda_{c}^{m}}{4!}  &  =2.9\pm0.2,\ \nu=1.3\pm0.1,\ a=0.36\pm
0.06,\text{ }\tilde{\chi}_{8}^{2}=1.1\ \text{(partial),}\nonumber\\
\ \tfrac{\lambda_{c}^{\left\langle \phi\right\rangle }}{4!}  &  =2.5\pm
0.1,\ \beta=0.11\pm0.02,\ b=0.65\pm0.04,\ \tilde{\chi}_{8}^{2}=0.88.\nonumber
\end{align}

\noindent These results are subject to errors due to the finite size $L$ and
finite cutoff scale $\Lambda$. We will use our data for different values of
$L$ and $\Lambda$ to extrapolate to the limit $L\rightarrow\infty,$
$\Lambda\rightarrow\infty.$ For the parameters $\nu$, $a$, $\beta$, $b$ we use
the naive asymptotic form%
\begin{equation}
x(L,\Lambda)=x+\tfrac{1}{\Lambda^{2}}x_{\Lambda^{2}}+\tfrac{1}{L^{2}}x_{L^{2}%
}+\cdots.
\end{equation}
For the critical couplings $\lambda_{c}^{m}$ and $\lambda_{c}^{\left\langle
\phi\right\rangle }$, however, we modify the finite $L$ correction according
to the finite-size scaling hypothesis \cite{fisher}%
\begin{equation}
\lambda(L,\Lambda)=\lambda+\tfrac{1}{\Lambda^{2}}\lambda_{\Lambda^{2}}%
+\tfrac{1}{\left|  L\right|  ^{1/\nu}}\lambda_{\left|  L\right|  ^{1/\nu}%
}+\cdots.
\end{equation}
The results we find are\footnote{Due to the relatively weak dependence on $L$
and $\Lambda$, the reduced chi-squared values for our extrapolation fits are
quite small ($\ll1$) and do not provide a useful statistical measure.}%

\begin{align}
\tfrac{\lambda_{c}^{m}}{4!}  &  =2.5\pm0.2\pm0.1,\ \nu=1.3\pm0.2\pm
0.1,\ a=0.43\pm0.05\pm0.02\text{ (all)},\\
\tfrac{\lambda_{c}^{m}}{4!}  &  =2.3\pm0.2\pm0.1,\ \nu=1.4\pm0.2\pm
0.1,\ a=0.48\pm0.06\pm0.02\text{ (partial)},\nonumber\\
\ \tfrac{\lambda_{c}^{\left\langle \phi\right\rangle }}{4!}  &  =2.5\pm
0.1\pm0.1,\ \beta=0.13\pm0.02\pm0.01,\ b=0.71\pm0.04\pm0.03.\nonumber
\end{align}
The first error bounds include inaccuracies due to Monte Carlo statistics,
higher energy states (for the mass curves), and extrapolation. The second
error bounds represent estimates of the systematic errors due to our choice of
initial state, time step parameter, and bin sizes in the DMC simulations.
\ For data generated from the mass curves, the main source of error was due to
extrapolation. \ The extrapolation error for the vacuum expectation value data
was still the most significant, though considerably smaller than that for the
mass calculations. \ This reduction is probably a result of the method used to
measure the vacuum expectation value.\footnote{We are refering to the result
$\left\langle 0^{+}\right|  \phi\left|  0^{+}\right\rangle =z_{\max}$, which
eliminates peak broadening effects for finite $L$. \ This is discussed at the
end of the previous section.} \ Our results for the critical exponents are
consistent with the Ising model predictions%
\begin{equation}
\nu=1,\quad\beta=\tfrac{1}{8}.
\end{equation}
The results for the critical coupling $\tfrac{\lambda_{c}^{m}}{4!}$ and
$\tfrac{\lambda_{c}^{\left\langle \phi\right\rangle }}{4!}$ are in agreement
with the recently obtained lattice result \cite{loinaz}\footnote{Critical
exponents were not measured in this study.}%
\begin{equation}
\tfrac{\lambda_{c}}{4!}=2.56_{-.01}^{+.02}.
\end{equation}

\section{Summary}

We have discussed the generalization of spherical field theory to other modal
expansion methods, in particular, periodic field theory. \ Using periodic
field methods we have analyzed two-dimensional $\phi^{4}$ theory and computed
the critical coupling and critical exponents $\nu$ and $\beta$ associated with
spontaneous breaking of $\phi\rightarrow-\phi$ reflection symmetry. \ Our
value of the critical coupling is in agreement with a recent lattice
calculation, and our values for the critical exponents are consistent with the
critical exponents of the two-dimensional Ising model. \ This lends support to
the popular belief that the two theories belong to the same universality class.

The full set of diffusion Monte Carlo computations used in our analysis
required about 30 hours on a 350 MHz\ PC processor. \ Complete codes can be
obtained upon request from the authors. \ The required computational time
appears to be dominated by the number of operations required to update the
Hamiltonian, which scales as $N_{\max}^{2}$.\ \ Errors can be reduced quite
substantially by using larger values of $L$ and $N_{\max}$ and utilizing
large-scale parallel processing.\ \ No less important, however, is that
periodic field theory provides a simple and efficient approach to studying
non-perturbative phenomena with only modest computer resources. \ Improvements
are now under way to utilize fast Fourier transform methods \cite{kogut} and
increase the computational speed. Future studies have been planned to analyze
phase transitions in other field theory models.{\normalsize \bigskip}

\noindent{\Large \textbf{Acknowledgment}}{\normalsize \bigskip}

We are grateful to Eugene Golowich for useful advice and discussions. We also
thank Jon Machta for comments on finite-size scaling and the referee of the
original draft for suggesting several improvements. Support provided by the
National Science Foundation under Grant 5-22698.{\normalsize \bigskip}

\noindent{\Large \textbf{Figures}}\bigskip

\noindent Figure 1. The only divergent diagram, which can be cancelled by
normal ordering.

\noindent Figure 2. Plot of $g(z)$ for $L=2.5\pi$ and $L=5\pi$. \ In each case
$\tfrac{\lambda}{4!}=2.76$ and $\Lambda=4$ .

\noindent Figure 3. Plot of $m$ as a function of $\frac{\lambda}{4!}$ for
$L=2.5\pi,$ $N_{\max}=10$.

\noindent Figure 4. Plot of $\left\langle 0^{+}\right|  \phi\left|
0^{+}\right\rangle $ as a function of $\frac{\lambda}{4!}$ for $L=2.5\pi,$
$N_{\max}=10$.

\newpage

\begin{figure}[ptb]
\epsfbox{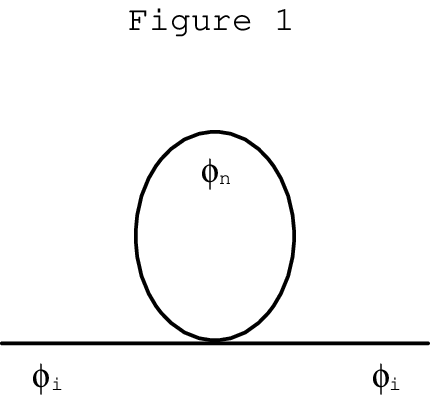}\end{figure}\begin{figure}[ptbptb]
\epsfbox{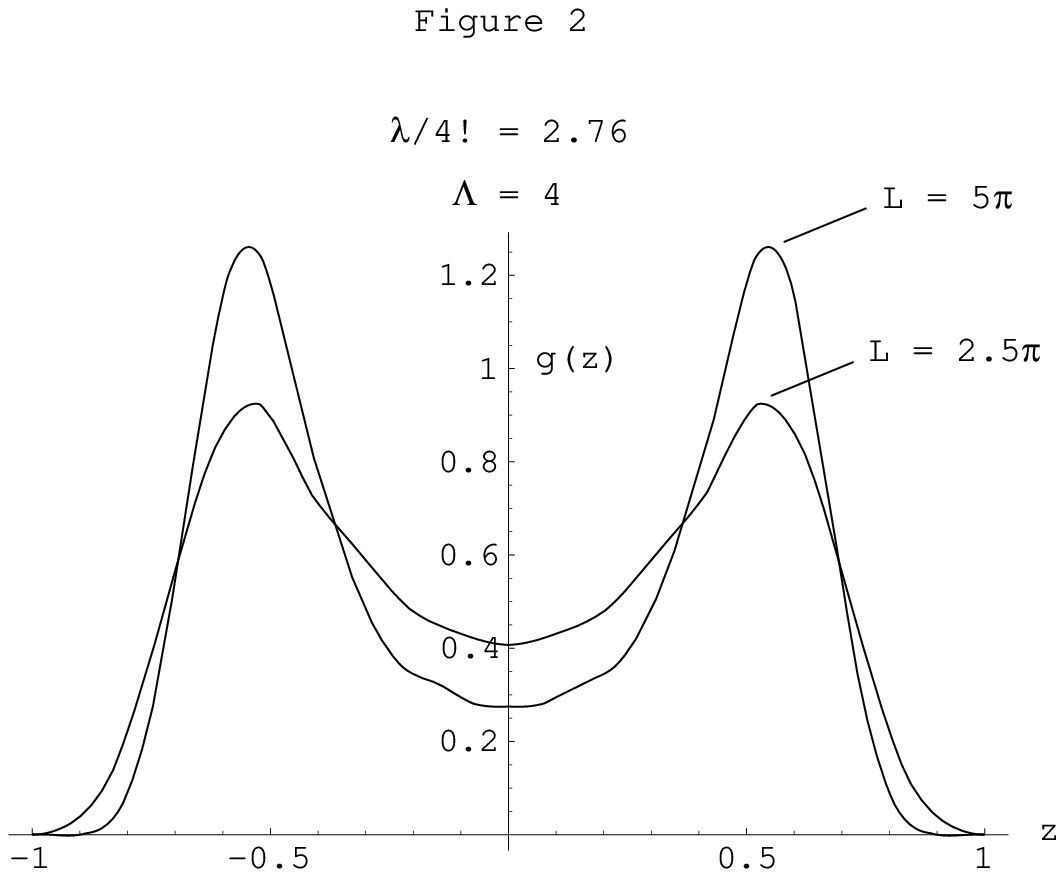}\end{figure}\begin{figure}[ptbptbptb]
\epsfbox{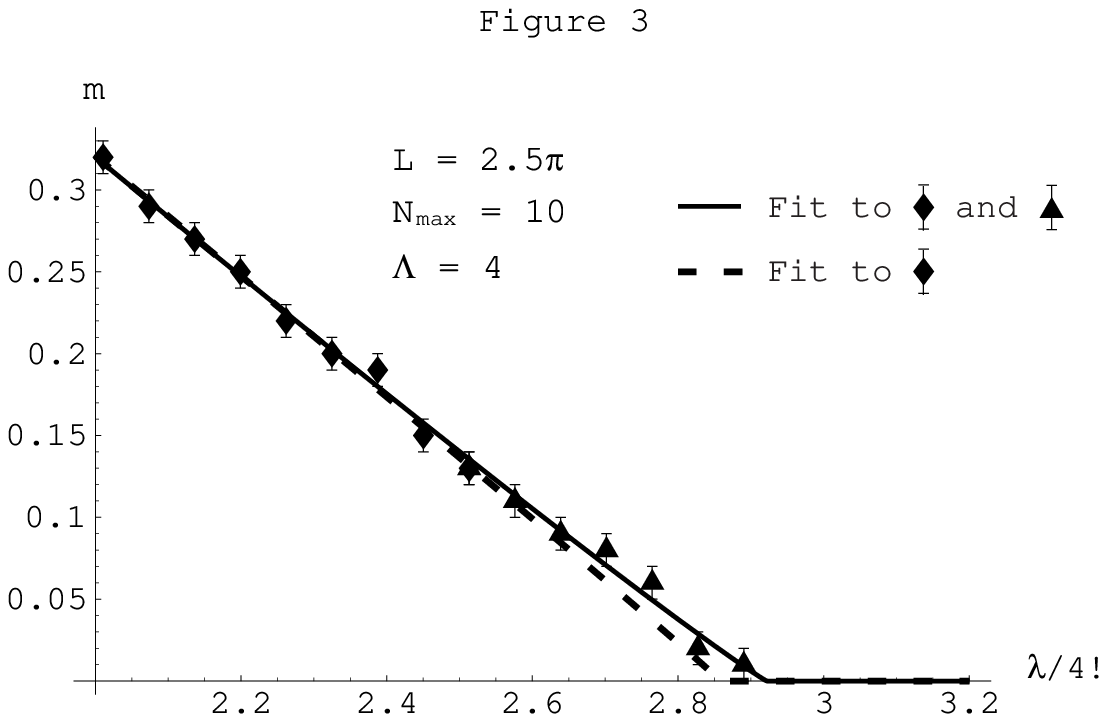}\end{figure}\begin{figure}[ptbptbptbptb]
\epsfbox{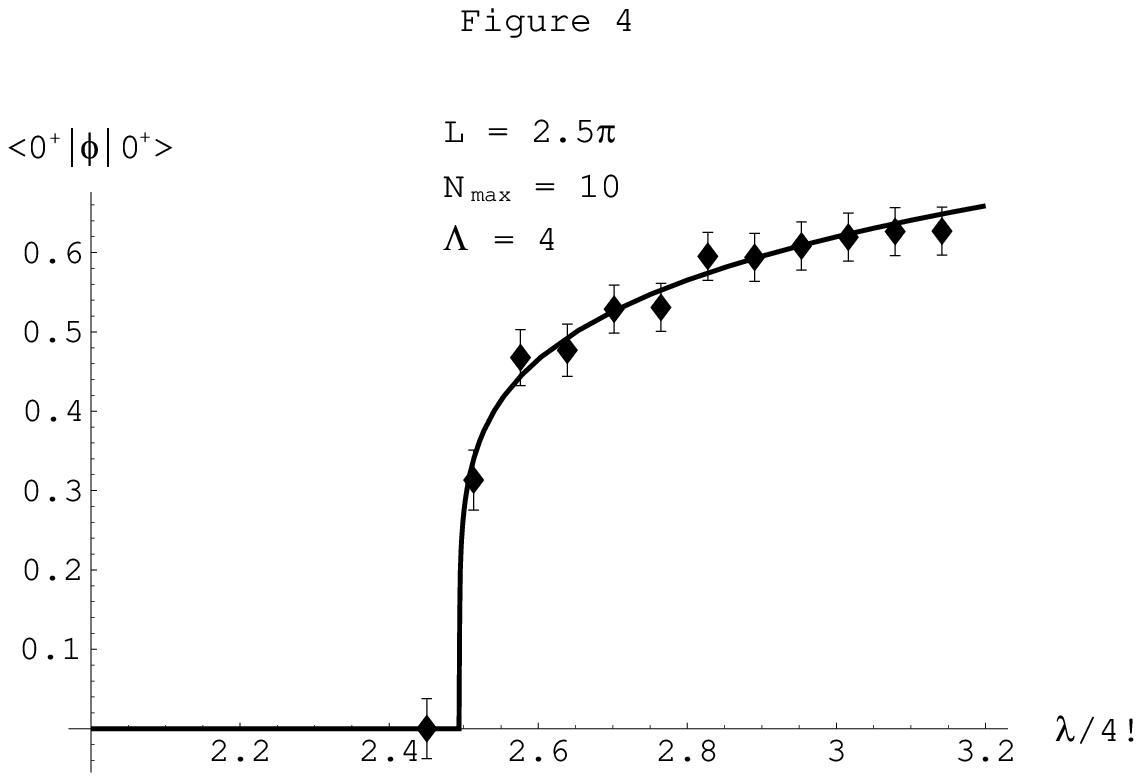}\end{figure}
\end{document}